\documentclass[twoside,twocolumn,10pt]{article}

\usepackage{wscg}           
\RequirePackage{ifpdf}
\ifpdf
 \RequirePackage[pdftex]{graphicx}
 \RequirePackage[pdftex]{color}
\else
 \RequirePackage[dvips,draft]{graphicx}
 \RequirePackage[dvips]{color}
\fi

\usepackage{nopageno}       

\newcommand\tab[1][1cm]{\hspace*{#1}}

\title{Mitigation of Social Media Platforms Impact on the Users }

\author{
\parbox{0.25\textwidth}{\centering
Smita Khapre\\[1mm]
Department of Computer Science \\
University of Colorado Colorado Springs\\
Colorado Springs, CO 80908, USA  \\[1mm]
skhapre@uccs.edu
}
\hspace{0.15\textwidth}
\parbox{0.25\textwidth}{\centering
Sudhanshu Semwal\\[1mm]
Department of Computer Science \\
University of Colorado Colorado Springs\\
Colorado Springs, CO 80908, USA  \\[1mm]
ssemwal@uccs.edu
}
\hspace{0.05\textwidth}
\parbox{0.25\textwidth}{\centering
}
}

\usepackage{url}
\makeatletter
\def\Uslash{\mathbin{\mathchar`\/}\@ifnextchar{/}{\kern-.15em}{}}
\g@addto@macro\UrlSpecials{\do \/ {\Uslash}}
\def\Ucolon{\mathbin{\mathchar`:}\@ifnextchar{/}{\kern-.1em}{}}
\g@addto@macro\UrlSpecials{\do : {\Ucolon}}
\makeatother

\begin{document}

\twocolumn[{\csname @twocolumnfalse\endcsname

\maketitle  

\begin{abstract}
\noindent
Social Media Platforms offer numerous benefits and allow people to come together for various causes. Many communities, academia, government agencies, institutions, healthcare,  entertainment, and businesses are on Social Media Platforms. They are intuitive and free for the users. It has become unimaginable to have life without social media. Their architecture and data handling are geared towards scalability, uninterrupted availability, and both personal and collaborative revenue generation. Primarily, artificial intelligence algorithms are employed on stored user data for optimization and feeds. This has the potential to impact their users' safety, privacy, and security, even if metadata is used. A new decentralized data arrangement framework based on the Fractal-tree and L-Systems algorithm is proposed to mitigate some of the impacts of Social Media Platforms. 

Future work will focus on demonstrating the effectiveness of a new decentralized framework based on Fractal trees and L-systems by comparing its results against state-of-the-art security methods currently employed in databases. The cryptographic algorithm could also be implemented for our framework, employing a new key generation for each branch. This will strengthen the database security, for example, when the user key is leaked, the change of each branch key will keep the data secure by employing a defense in the proposed framework of L-System's tree.  
\\

\end{abstract}

\subsection*{Keywords}
L-System, Fractal, Koch Curve, Social Media Platform, User Engagement, Context-sensitive, Parametric, Stochastic, Data Arrangement, OSINT, Script-kiddie, Reconnaissance, Non-repudiation, Elasticity, Scalability, Object Oriented Programming, and Processing. 
\vspace*{1.0\baselineskip}
}]


\section{Introduction}

\copyrightspace

Social Media Platforms (SMP) like Facebook, Instagram, Twitter, LinkedIn, YouTube, Discord, and TikTok are very popular among youth and adults. They are the cheapest source of knowledge, information, and entertainment. They connect people globally. The user data mainly the private data, shares, and posts on SMP can be accessed by businesses, algorithms, users, and adversaries on SMP. This is called Open-source Intelligence (OSINT) \cite{HAckingtheHuman}. Anyone can do it as it is publicly available. These are not just used by government and adversaries, but even a novice known as script-kiddie \cite{ScriptKiddieOrNationStates_gaining}\cite{ScriptKiddie} in cybersecurity. They are used for reconnaissance \cite{Recoinnassance} by various agencies, adversaries, users, and businesses. Privacy settings are available to curb data theft and leaks to a certain extent. Once the user data is on social media, it is always accessible to other user accounts. Also, the privacy settings are not intuitive and difficult for most users. The SMP algorithms, deployed for revenue generation by SMP providers to feed sponsored content, use a huge amount of user data with user consent. The user is unaware of the Artificial Intelligence (AI) algorithm and how it manipulates users' psyche. The availability of user data publicly causes various impacts \cite{smp_UE_shahbaznezhad2021role} on users' Safety, Privacy, and Security (SPS) \cite{AI_SFS} as they are not just being used for good cause but illegal trades as well. The information on the SMP cannot be assumed to have a trusted source. The SMPs are not governed by law. There is a lot of misinformation and fake news on SMPs which causes societal instability and impacts the mental and physical well-being of users. One such example is The Capitol Attack on Jan 6, 2020. The recent Senate hearing on STOP CSAM Act \cite{CSAM} is an eye-opener.

To mitigate these SMP impacts, one must understand the issues it poses. The major concerns of the SMP are their opacity, absence of regulation, and lack of liability.
\begin{itemize}
    \item How is user data organized and stored on SMP? Who can access the user data, and what is the purpose of their access?
    \item What kind of AI algorithm implementation on user data is prevalent on SMP? How do they impact the user?
    \item Which interactions are allowed between the AI algorithms and the user on SMP?
    \item What is the data control mechanism?
    \item Are there any report generation on user activities, algorithms' user data utilization, or other accounts' access to the user data on demand? \\
\end{itemize}

To deal with these issues, we think a paradigm shift is warranted. Right to be Forgotten \cite{alfaidi2022right} is required for the user on SMP. Every comment, like, and share might not be required to be stored in SMP DB. Similarly, every feed a user scrolls through, may not characterize the user for recommendation.
The main paradigm shift which we are proposing is that the user should have control of their data, with legal rules in place if necessary. This will also allows the SMP posts to originate from a trusted source when public viewing of the data is allowed by the owner as well as the SMP platforms.  The main paradigm shift is that SMPs will not own the data or the meta data which might disrupt SMP's existence. It may possibly cause backlash to allow legal protection of those assets' usage by their own clients.  We believe that the users should retain some control over how their data is used, while SMP may be paid for managing this data without actually owning it, even the meta data \cite{JaromeLanierYouTubeHere}.
User/post/feeds with numerous followers/viewers should abide by journalism ethics of Truth, Accuracy, and Objectivity \cite{ward2019journalism}. 

We feel that our reasoning is holistic and have organic approach to mitigate some SMP impacts. It proposes that all the user information/data should be arranged in a tree structure (Fractals \cite{mandelbrot1982fractal} and Koch Curve) by L-System Algorithms \cite{prusinkiewicz2012algorithmic} providing the following benefits.
\begin{itemize}
    \item Removal of the data branch with untrusted source by the governing bodies. It should also remove their respective comments/shares/views.
    \item User can delete any data branch at their discretion which will delete the following shares, comments, and views. 
    \item Platform owner can delete compromised data branches thereby wiping off all their instances at once. 
    \item Non-repudiation \cite{non-repudiation}.
    \item Elasticity and Scalability \cite{elasticity-scalability}.
    \item Tracking the parameter of the number of likes, views, and shares for the Posts and Feeds. \\ 
\end{itemize}

\begin{figure*}[ht!]
    \centering
    \includegraphics[width=16 cm,height=7cm]{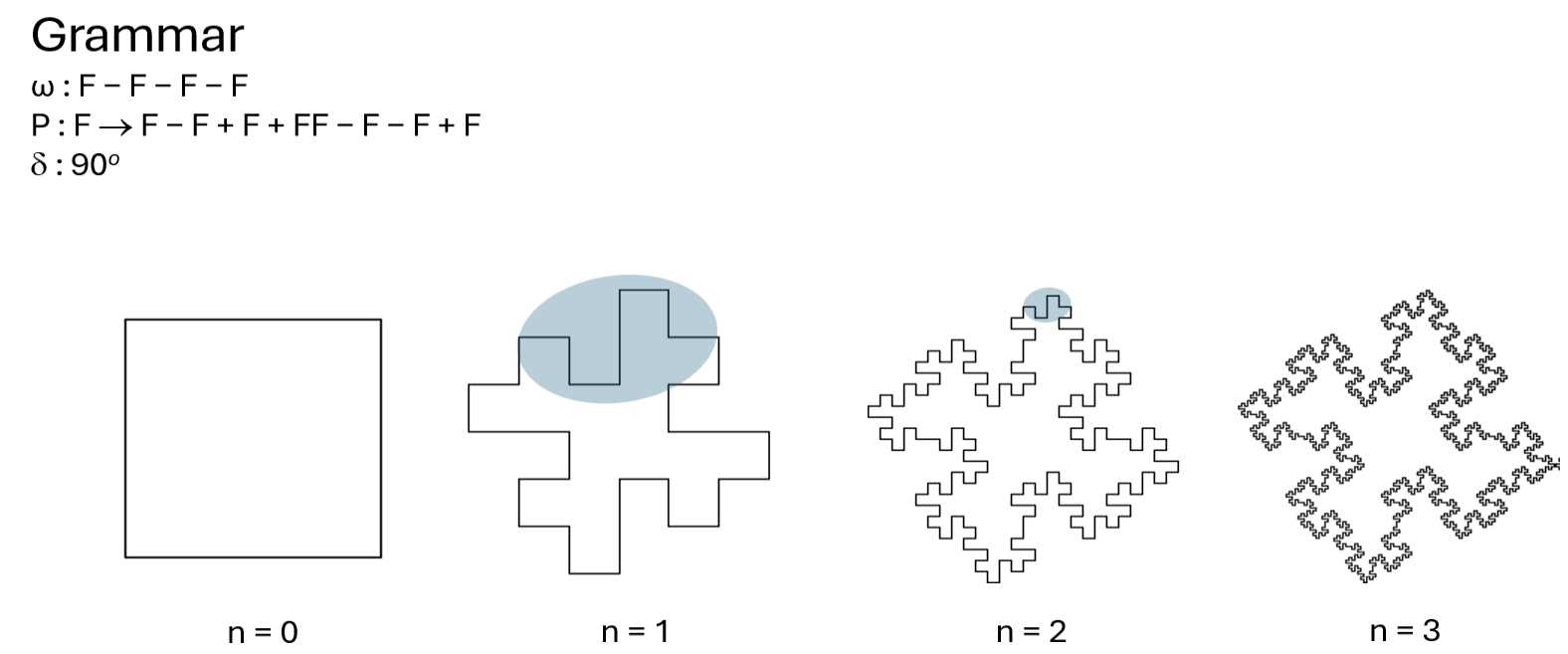}
    \caption{Quadractic Koch Curve Production}
    \label{fig:KochCurve}
\end{figure*}

Feed is not a form of user data, thus it must be dealt with differently. The Feed parameters of likes, views, and shares, play an important role in forming the user's character. An AI algorithm uses it to recommend the next feed to the user. It also plays with user psychology. It influences user friends too, by feeding the user feeds activity as the feeds and notification to friends account.

The proposed project intends to solve some of the SMP impacts by presenting the design of the data arrangement. The organic structure of the trees is exploited using L-systems \cite{prusinkiewicz2012algorithmic}. The L-Systems were introduced by Lindenmayer in 1968 to explain the theoretical framework of the development of multicellular organisms, which is later applied to plants. The geometric features implementation to model plants using L-Systems allowed computer graphics to visualize plant structures realistically. It also represents their developmental processes. L-systems algorithm is a part of formal language theory. Explained in the theory of algorithms \cite{prusinkiewicz2012algorithmic}. 

The Fractals \cite{mandelbrot1982fractal} are the self-similar structures generated by the L-System algorithm. It has finite features leading to infinite self-similar structures, thereby rendering a perfect pattern. L-system \cite{prusinkiewicz2012algorithmic} is an algorithmic approach to generating a fractal. 
L-Systems can generate different types of fractals depending on grammar $G$, it may be stochastic, context-sensitive, parametric, and bidirectional. These are deterministic algorithms, starting at the left-most variable to apply the set of rules.

Koch \cite{von1906methode} and Mandelbrot \cite{mandelbrot1982fractal} gave a simple explanation of rewriting rules in the L-system algorithms. The grammar of rewriting rules is summarized as $G = (V, \omega, P)$, where production starts at Axiom $\omega$ (apex in trees), has a set of variables V, and a set of rules P to translate each variable based on rules with every iteration. It is a Deterministic Algorithm, which starts from the leftmost variable and applies a set of rules.
The number of times the P is applied is called iterations $n$. The Koch Curve uses the angle of movement and is called Turtle Interpretation. Figure \ref{fig:KochCurve} shows the quadratic Koch Curve production. The ease of modifying L-systems makes them suitable for developing new Koch curves. For example, one can start from a particular L-system and observe the results of inserting, deleting, or replacing some symbols. 

\section{Approach}
Assuming the user account as a tree trunk, the branches as posts and followers, and their leaves as the shares, comments, views, and likes; it allows efficient data arrangement and centralized management. It ensures that the user controls its own data and access to other entities. The proposed L-system algorithms generate the fractals inspired by the organic plant structure.
\begin{figure}[ht!]
    \centering
    \includegraphics[width=7cm,height=5cm]{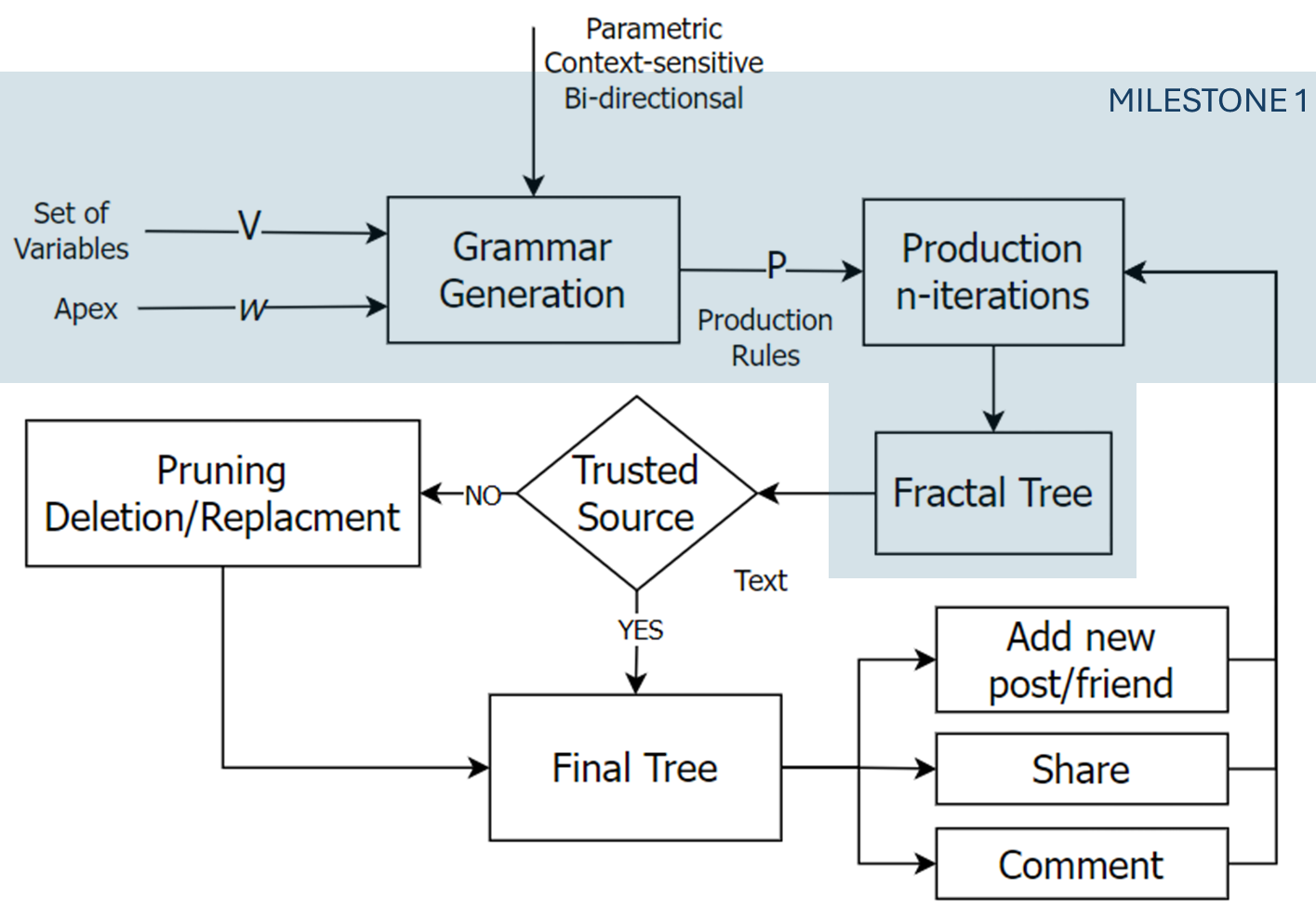}
    \caption{Proposed Design}
    \label{fig:Block}
\end{figure}
\begin{figure*}[ht]
    \centering
    \includegraphics[width=14.5cm,height=8cm]{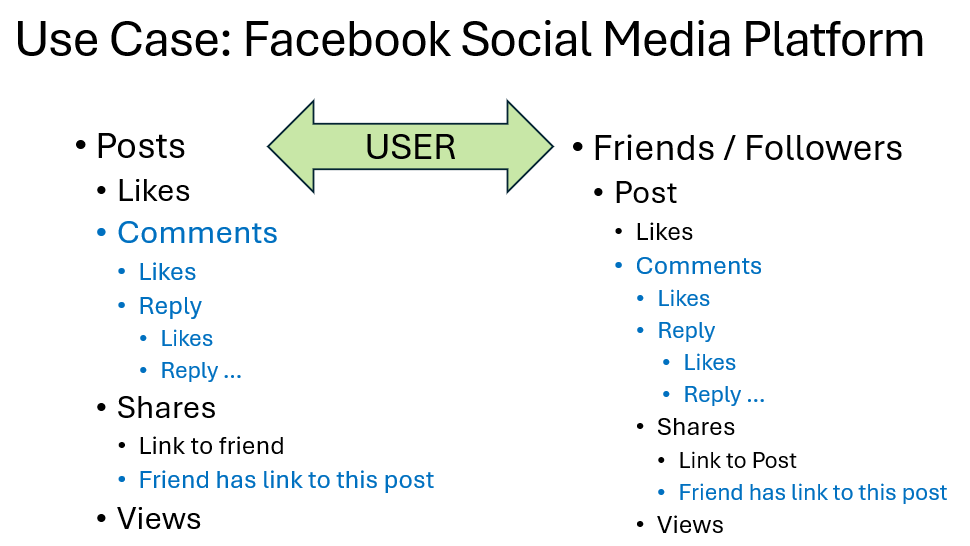}
    \caption{Usecase: Facebook}
    \label{fig:FB}
\end{figure*}
\begin{figure*}[ht!]
    \centering
    \includegraphics[width=14.5cm,height=9cm]{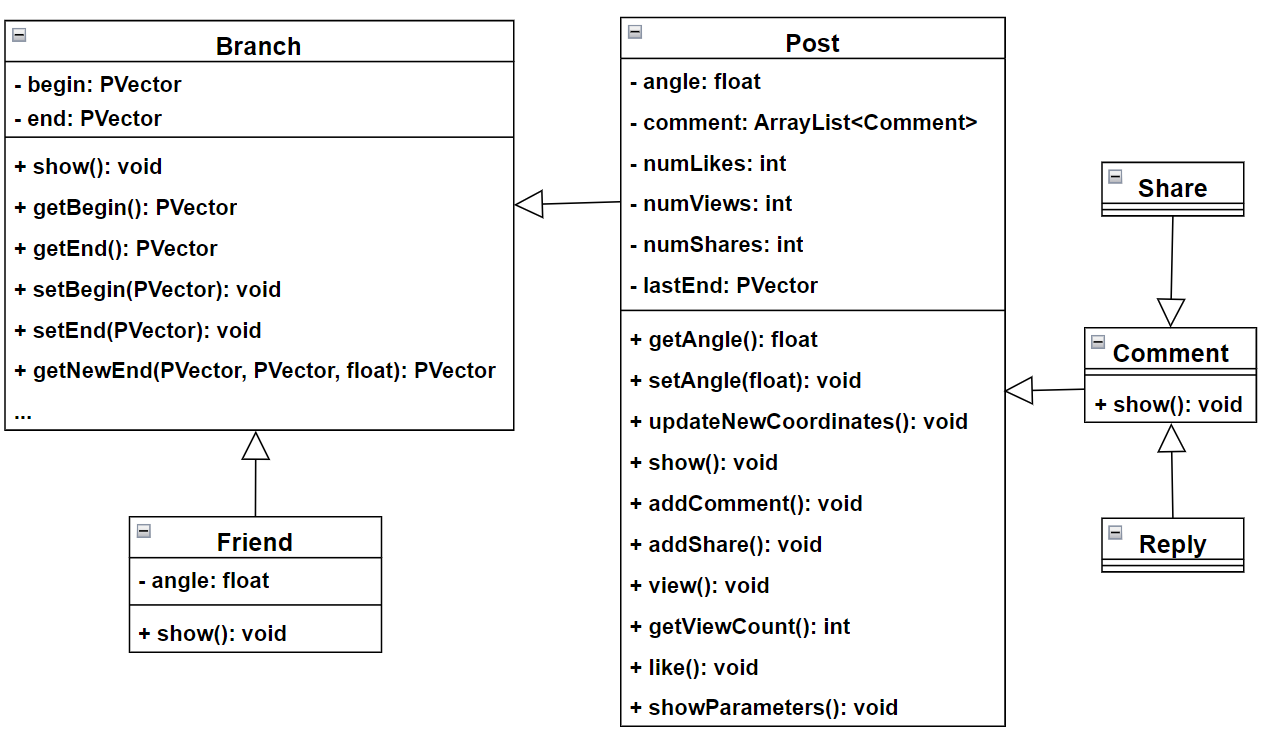}
    \caption{Class Diagram of OOP Design}
    \label{fig:class}
\end{figure*}

The proposed design in Figure \ref{fig:Block} allows the non-repudiation of the post. The trusted source validation can be done at the source, for misinformation and fake news. If the source is untrustworthy, it can be restricted from being accessed by other users on the SMP. The Facebook SMP is assumed as a use-case for this project as shown in Figure \ref{fig:FB}. Each user has either posts or friends. Each post has likes, comments, shares, and views. Each friend has posts. The object-oriented design is proposed in Figure \ref{fig:class}.
The key assumptions in the proposed design are listed below.
\begin{itemize}
    \item Data is encrypted at rest and in transit.
    \item Each class like user Branch, Post, Friend, Comment, Share, and Reply will have a data element to store the respective data.
    \item It is an event-based framework architecture and is updated with an event like addition of a post, addition of a friend, comment on the post, sharing a post, like a post, and view a post.
    \item To avoid the complexity in understanding, the visualization is limited to 2-dimensions. 
    \item The feeds are not in the scope of the project. 
\end{itemize}

The data arrangement in fractal-plant structure with context-sensitive, bidirectional, stochastic, and parametric grammar is proposed. It has branches linked to the post and related information like \textit{comments, shares, views, and likes}. At any point, the user or the governing body may choose to delete a particular post by deleting the entire branch without affecting other posts or followers. The grammar in this project is given by 
\begin{equation}
    G = (V, \sum, \omega, \textbf{P})   
\end{equation}
Production starts at Axiom '$\omega$' (apex in trees), has a set of variables '$V$', a set of parameters '$\sum$', and a set of rules '$\textbf{P}$' to translate each variable based on rules with every iteration. 

$\omega = \{ U \} $ 

$V = \{ P, F, C, S, r, l, m, k, [, ], \}$

$\sum = \{ i, s, v \}$ \\

$\textbf{P} = \{$ 
\tab $\omega \rightarrow mmmmmPF$, \\
\tab \tab $F \rightarrow mm[[rmk]F]$, \\
\tab \tab $P \rightarrow mmm[[lBCS]P]$, \\
\tab \tab $C \rightarrow m[rmmk]C$, \\
\tab \tab $S \rightarrow m[lmmk]S$, \\
\tab \tab $B \rightarrow BB$ \\
\tab $\}$\\

'$U$' represents the user trunk and apex of the fractal plant, '$P$' represents the posts, '$F\textbf{}$' represents the followers, '$C$' represents the comments, '$S$' represents the shares of the post by other users or followers, '$r$' represents the right turn and '$l$' represents the left turn, '$[$' represents the object beginning for bidirectional grammar, '$]$' represents the return to previous object for bidirectional grammar, '$i$' represents the number of likes for the post, '$s$' represents the number of shares for the post, and '$v$' represents the number of views for the post. $\textbf{P}$ is the set of production rules where $m$, $l$, $r$, and $k$ are the constants and do not have replacement rules for growth. 

\begin{figure}[ht!]
    \centering
    \includegraphics[width=7cm,height=10cm]{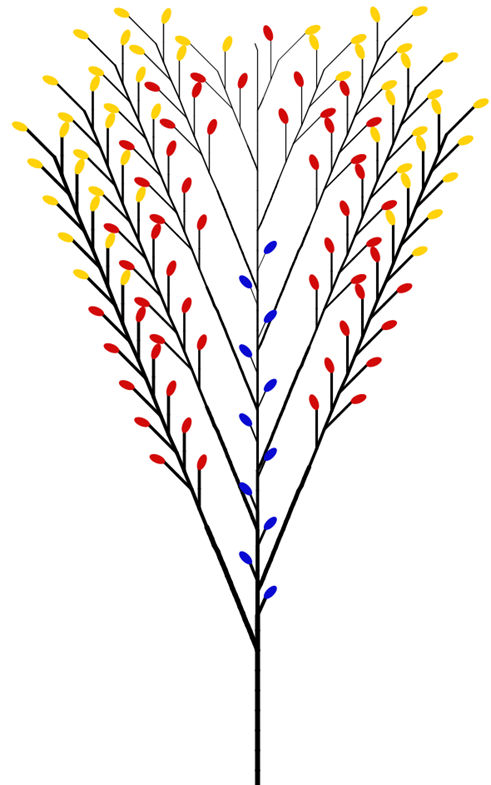}
    \caption{Generated Tree with Context-Free Grammar}
    \label{fig:Tree}
\end{figure}

\begin{figure}[ht!]
    \centering
    \includegraphics[width=7.7cm,height=8cm]{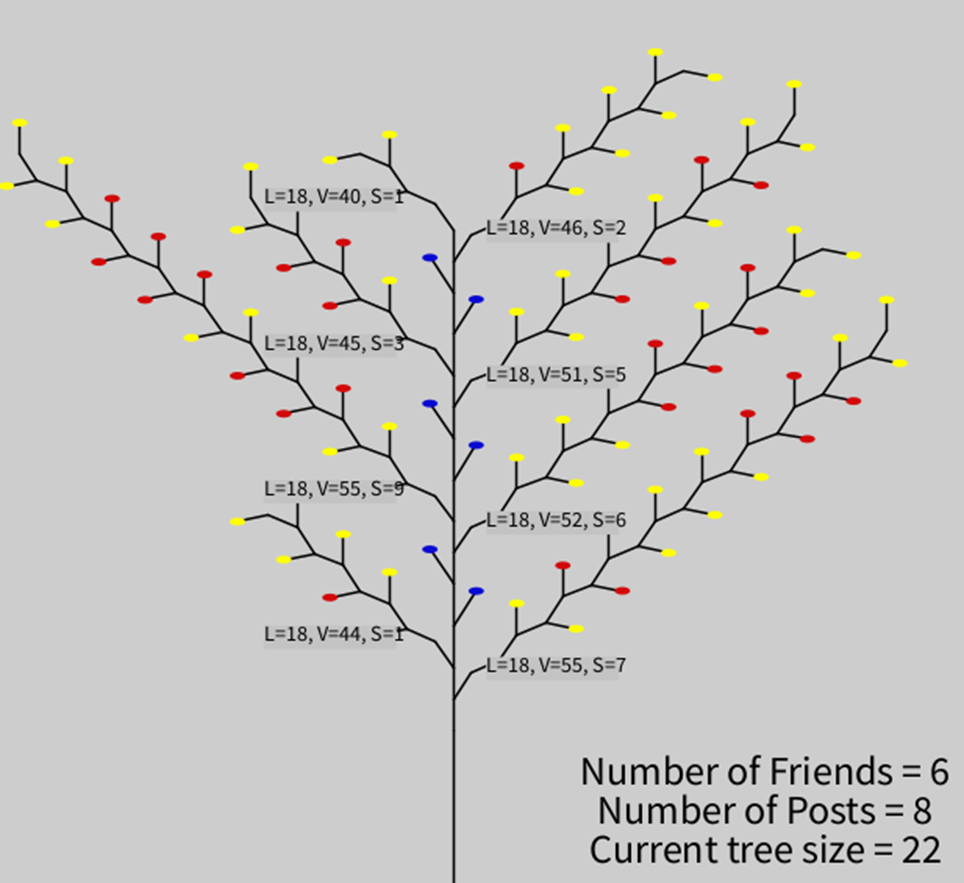}
    \caption{Final Tree with Context-Sensitive Grammar}
    \label{fig:OOP-Tree}
\end{figure}

\begin{figure*}[ht!]
    \centering
    \includegraphics[scale=0.87]{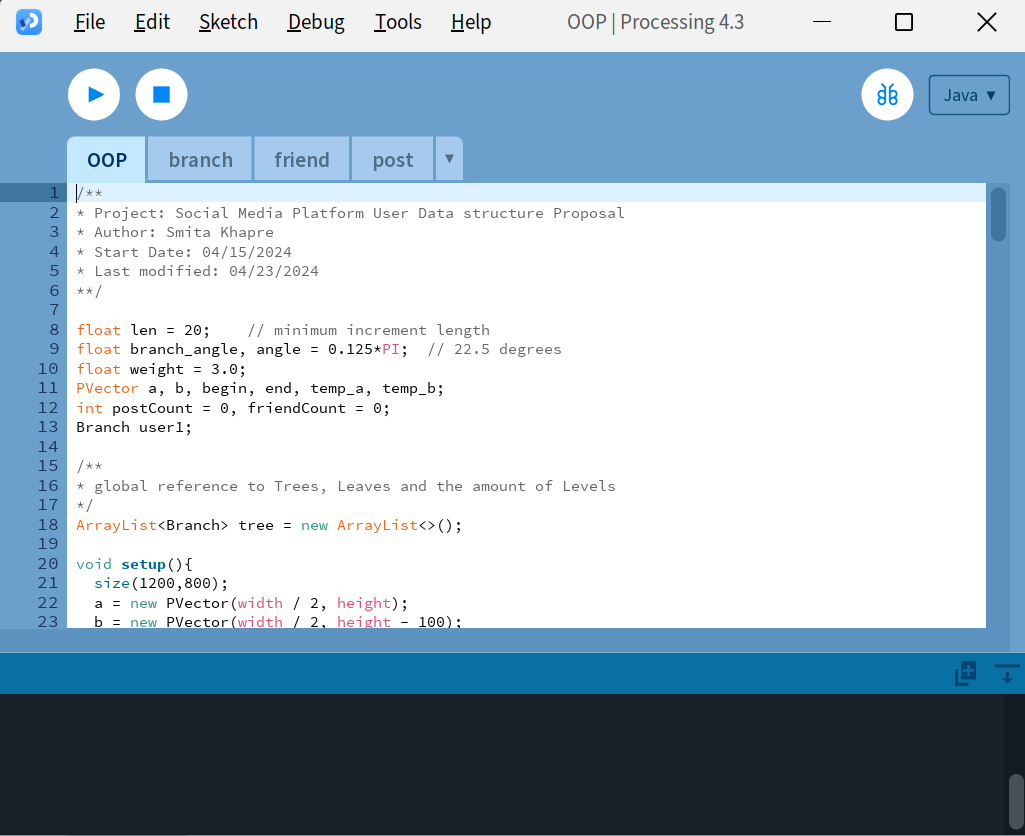}
    \caption{Processing Application}
    \label{fig:Processing}
\end{figure*}
Figure \ref{fig:Tree}, shows the tree of the user account with the mentioned grammar as a context-free L-system algorithm for milestone 1. The context-free grammar generates a symmetrical tree structure. It is implemented to evaluate the correctness of the production rules. The blue-colored leaves are the friends or followers with the link to the user account, the red-colored leaves are the comments on the post, and the yellow-colored leaves are the shares of the post. The context-sensitive and parametric grammar is shown in Figure \ref{fig:OOP-Tree} is implemented as per the designed algorithm in section \textit{Algorithm Design}, as milestone 2. It uses Object Oriented Programming (OOP) Design concepts in Java. 

We used Processing 4.3 to implement our ideas. Processing is a flexible software sketchbook. It supports Java, JavaScript, and Python programming languages on the operating systems of Windows, Linux, and MacOS. For this project, Java programming language on the Windows machine is used. The executable generated can be executed on Windows (64-bit) machines with JRE. On machines other than Windows, it requires OpenJDK 17 installation.

In the Processing 4.3 application as shown in Figure \ref{fig:Processing}, there are four tabs. The first tab is \textbf{OOP}. It is the sketchbook, which contains initialization code, \textit{setup()} and \textit{draw} routines. The second tab is \textbf{Branch}. It contains the parent class \textbf{Branch} definition. The third tab is \textbf{Friend}. It contains the sub-class \textbf{Friend} definition. And the fourth tab is  \textbf{Post}. It contains the sub-classes \textbf{Post, Comment} and \textbf{Share} definitions.

\section{Algorithm Design}

In Processing sketchbook, there are two main routines. One is Setup and the other is Draw. Setup routing sets up the screen dimensions and initial parameters. It initializes the environment. Draw routine continuously runs after the setup routine based on the frame rate. The lowest possible frame rate is 1. 

Event-based routines are mouse-driven or keyboard-driven. Mouse-driven works on the mouse pointer coordinates on the screen. It uses either the mouse movements or the mouse clicks. Keyboard-driven uses the key presses. For contest-sensitive L-System design, event-based programming is done with key presses as it has seven events. 

Firstly, define the Branch, Friend, Post, Comment, and Share classes based on the design in the class diagram shown in Figure \ref{fig:class}. 
The Branch class is the parent class. The two classes Post and Friend are inherited from the Branch parent class. The class Comment is inherited from the Post class, And lastly, the class Share is inherited from the Comment class.\\

In the \textit{setup()} routine, 
\begin{itemize}
    \item Initialize length of branch segment, branch angle, twig angle, post count and friend count
    \item Initialize the screen width and height
    \item Instantiate the user apex of the tree as the Branch object at the bottom middle of the screen \\
\end{itemize}

In the \textit{draw()} routine, 
\begin{itemize}
    \item Initialize background and stroke parameters.
    \item Draw the Apex of the tree.
    \item Check for tree branches.
    \begin{itemize}
        \item Draw the existing branches. \\
    \end{itemize}
\end{itemize}

In the \textit{keyPressed()} routine,
\begin{itemize}
    \item  If key 'p' or 'P' is pressed, then add a Post object as a branch to the tree.
    \item  If key 'f' or 'F' is pressed, then add a Friend object as a branch to the tree.
    \item  If key 'c' or 'C' is pressed, then get a random post index and add the Comment object to the Post's Comment object ArrayList.
    \item  If key 's' or 'S' is pressed, then get a random post index and add the Share object to the Comment's object ArrayList.
    \item  If key 'l' or 'L' is pressed, then increment like count of all Post objects in the tree.
    \item  If key 'v' or 'V' is pressed, then increment like count of all Post objects in the tree.
    \item  If key 'r' or 'R' is pressed, then check if the number of views is greater than 50 for all Post objects in the tree, 
    \begin{itemize}
        \item If Yes, remove the post object.
        \item If No, continue. \\
    \end{itemize}	
\end{itemize}

The event of pressing the keys 'C' and 'S' mimics the event of another user commenting on and sharing the post respectively. It also increments the number of views parameter for the respective post. The events of pressing the keys 'L' and 'V' mimics the events of another user likes, and views the post. Like event also increments the number of views parameter.

There is a common routine \textit{printStatus()}, which prints the number of branches, posts, and friends of the tree on the bottom of the screen.

In the class \textbf{Branch}, there are two private PVector members, 'begin' and 'end'. And, it has four public methods, getBegin, getEnd, setBegin, setEnd, getNewEnd, and show for getting, setting, and drawing the branch coordinates. These six methods are inherited by the sub-classes Post, Friend, Comment, and Share. 

\begin{itemize}
    \item \textit{getBegin} method returns the starting point coordinates of the PVector begin variable.
    \item \textit{getEnd} method returns the ending point coordinates of the PVector end variable.
    \item \textit{setBegin} method sets the starting point coordinates of the PVector begin variable using the input parameter.
    \item \textit{setEnd} method sets the ending point coordinates of the PVector end variable using the input parameter.
    \item \textit{getNewEnd} method takes the three input parameters, 2 PVector variables begin and end, and a float variable angle. It generates and returns a new PVector coordinate variable by changing the direction using the opposite angle variable. 
    \item \textit{show} method draws the branch from starting to ending coordinates. 
    \begin{itemize}
        \item In \textbf{Friend} class, \textit{show} method additionally draws a blue-colored ellipse to indicate the user link of the friend.
        \item In \textbf{Comment} class, \textit{show} method additionally draws a yellow-colored ellipse to indicate the user link of the friend who has commented on the respective post.
        \item In \textbf{Share} class, \textit{show} method additionally draws a red-colored ellipse to indicate the user link of the friend who has commented on the respective post.\\
    \end{itemize}
\end{itemize}

The other methods \textit{getAngle(), setAngle(), updateNewCordinates(), addComment(), addShare(), showParameters(), like(), view(), getViewCount(), and show()}, are overloaded in the \textbf{Post} class. These methods are responsible for the tree structure data arrangement and visualization. 

\begin{itemize}
    \item \textit{getAngle()} method returns the angle of the Post branch. 
    \item \textit{setAngle()} method takes the angle as an input parameter and sets it for the Post branch. 
    \item \textit{updateNewCordinates()} method calculates the new coordinate of the next branch and sets them as coordinates of begin and end variables of the post. 
    \item \textit{addComment()} method gets the current comment branch's coordinates, instantiate a new comment object, and adds it to the Comment ArrayList variable of the Post Object. It also increments the views parameter.
    \item \textit{addShare()} method gets the current comment branch's coordinates, instantiate a new Share object, and adds it to the Comment ArrayList variable of the Post Object. It also increments the views and shares parameters.
    \item \textit{showParameters()} method displays the likes, views, and shares parameters besides the post branch.
    \item \textit{like()} method increments the likes and views parameters.
    \item \textit{view()} method increments the views parameter.
    \item \textit{getViewCount()} method returns the views parameter.
    \item \textit{show()} draws the post branch, comments' branches, and shares' branches based on the number of comments and shares of the post respectively. It also displays the likes, views, and shares parameters beside each post branch. \\
\end{itemize}

\begin{figure}[ht!]
    \centering
    \includegraphics[width=7.7cm,height=8cm]{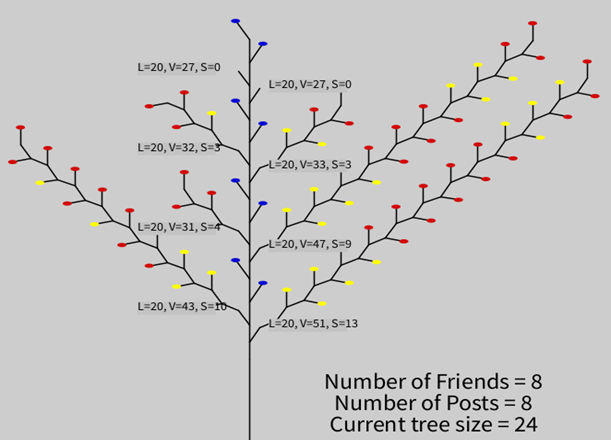}
    \caption{Step 1: Create a Fractal Tree}
    \label{fig:step1}
\end{figure}

\begin{figure}[ht!]
    \centering
    \includegraphics[width=7.7cm,height=8cm]{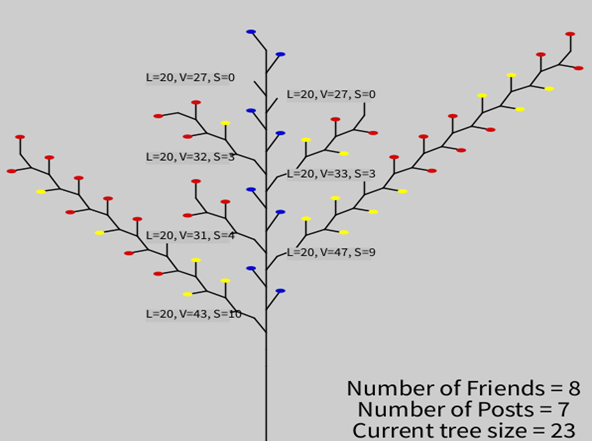}
    \caption{Step 2: Check if V exceeds 50, then Prune}
    \label{fig:step2}
\end{figure}

\begin{figure}[ht!]
    \centering
    \includegraphics[width=7.7cm,height=8cm]{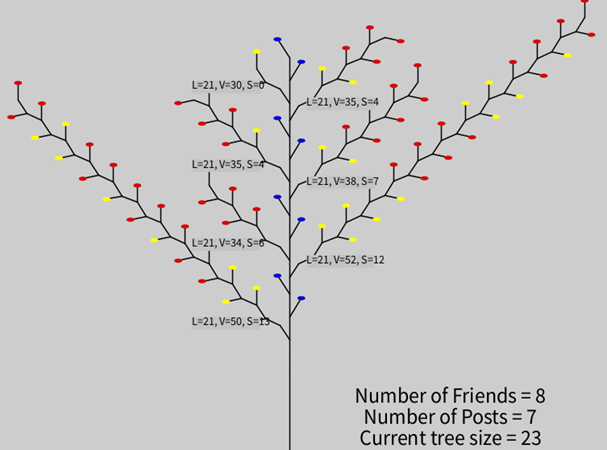}
    \caption{Step 3: Add more Comments, Shares, Likes, and Views}
    \label{fig:step3}
\end{figure}

\begin{figure}[ht!]
    \centering
    \includegraphics[width=7.7cm,height=8cm]{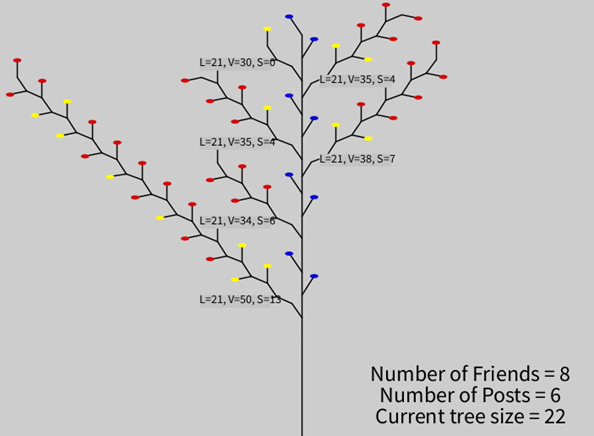}
    \caption{Step 4: Check if V exceeds 50, then Prune}
    \label{fig:step4}
\end{figure}

\section{Results}

Much work remains as we have only begun exploring the feasibility of our concept to visualize the L-system tree, with the potential to incorporate a wide range of information to support its growth. This represents our initial novel idea to begin the research, recognizing that much more work is required to achieve our goal. We need to compare the performance of state-of-the-art databases with our system to demonstrate the visualization aspects presented in this paper. We aim to provide performance metrics that will help validate aspects of our research.

The proposed project creates a fractal tree, an SMP data arrangement in the fractal tree structure; when the posts and friends are added using keypress 'P', and 'F' respectively. Also, when comments and shares are added by pressing the keys 'C' and 'S'. The parameters are displayed beside each post. At the bottom of the screen, on the right side of the tree trunk, it displays the summary of the generated fractal tree with the number of branch objects as the tree size, the number of the Friend objects, and the number of the Post objects. This is shown in Figure \ref{fig:step1}. At any given time key 'R' is pressed and if the views for any post exceeds 50, the corresponding post branch or branches are pruned. This is shown in Figure \ref{fig:step2}. Again generate more branches and pruning is shown in Figures \ref{fig:step3} and \ref{fig:step4} respectively. 

This framework allows the entire post and its comments, shares, and related parameters to be safely deleted without impacting other posts and users. It deletes all the references to the posts. None of the related history can be accessed. It also restricts the other users' access to the post if they are not in the friend list or shared with the public. Moreover sharing with the public will exceed the views threshold faster and will come under scrutiny. This will prevent fake news and misinformation to a certain extent. 

This framework is expected to be slower than conventional databases for data access as there is no direct way to access specific user data. Traversing through the apex would be the only way. While it adds security and an easier way to wipe the data, caching of data will add to the security challenge. The framework testing with real data, performance measurement, and comparison is out of scope for this project. 

\section{Comparison with current Database}
Facebook has a complex architecture \cite{facebookArchitecture} for data storage and caching. It uses Cassandra, MySQL, and HBase Databases, and Haystack for media file storage. To curb the latency, it uses Memcache for each user id, so that it does not need to access the Database for every query a service issues. This means that the user data is divided and stored in different stores.
Cassandra, MySQL, and HBase are column-based tabular and open-source databases. They all require database administrator, which poses a single point of failure threat.

The proposed framework could be beneficial and act as a defense-in-depth mechanism, enabling stronger encryption. Furthermore, it will prevent data theft and leakage. During the attack attempt on the individual Database, store, or even the memcache, there is no way to get the encryption keys. It is immune to Brute-force and Rainbow table attacks as the key for each post for each user is expected to be different. This proposed framework is a decentralized approach. Thereby, eliminating database administrator requirements and mitigating the single point of failure threat. 

\section{Future Work}
User's right to allow the visualization of the comments and replies on the post either by hiding or deleting them. This can be automated by having parameter design on comments for users to set or designed as event-based. The Trusted Source Evaluation for a post, when it reaches a threshold of views. Deep Learning technology can be explored to achieve it. SMPs reserve the right to Blacklist users if the number of posts deletions exceeds the threshold owing to untrusted sources and misinformation.
Handling the Feed data is an important part of SMP. It contributes to the revenue model. It probably has the most impact on society and public opinion. While SMPs focus on improving their revenues, they must be liable for the well-being of their society and forming public opinion. 
Deletion of all the posts or activities during a hacking event discovery. This is another important aspect. If we assume an ideal scenario, with no hacking activities, should all the posts, the history of parameters be retained by SMP at all times? What should happen to the user account and its data when someone dies? In the real organic world, every entity has a lifecycle, and every beginning has an end. Does it require timestamping for each post, so they age and die?

Report generation of the user activities and screen time on other SMP features like sponsored feeds, new feeds, and ads which is the main revenue generation source. The users get addicted or carried away when on SMP. A user has a right to know, how he/she is spending the time and what kind of content. Does it require SMP to categorize the feeds? What should be the basis of this categorization?

Database design using L-system algorithm-based fractal tree structures is a prospect. This moves away from traditional table-based to tree-based DB.
Could it eliminate the central Database Administrator requirement? The SMP, like Facebook with a colossal user base, the accessibility and availability of the services is a priority. Facebook utilizes memcache \cite{facebookArchitecture} to provide quick turnover. The feasibility of the proposed database design is another prospective work. 

The proposed project does not dwell on data at rest. 
There is a prospect where each branch would have different symmetric key cryptography. Though the L-system algorithms are deterministic, the production rule for each user could be different, and thus, a different key for each post. This enables stronger cryptography. Even if an adversary gains access to one key, only the corresponding branch can be decrypted. Other branches remain secure.

The role of Quantum Computing \cite{glassner2023QuantumComputing} cannot be neglected owing to recent developments. 
The current cryptographies are based on the time it takes to break a key, whether an RSA or an Elliptic Curve. By 2028, Quantum Computer could implement Shor's Algorithm, Grover's Algorithm, and Simon's Algorithm to break the existing cryptographies \cite{PQCsurvey}\cite{chaturvedi2024postQC}. All cryptographies will lose their defenses. There is a prospective requirement to design any privacy and security aspect with post-quantum cryptography \cite{chaturvedi2024postQC}. 
At the same time, quantum computer availability will enable the proposed design, to be implemented as a decentralized Database. And eliminate the current limitation of latency in accessing the data from the tree.

The suggested L-system based post quantum cryptography is another prospective future work and a promising new design. The grammar and the production rules can be used to determine the symmetric key for the individual branch. Currently, NIST Post Quantum Cryptography (PQC) Standardization \cite{PQCsurvey} is in the process of establishing standardized cryptographic methods showing resilience to quantum computing vulnerabilities. NIST PQC has selected CRYSTALS-KYBER, CRYSTALS-Dilithium, FALCON, and SPHINCS+ algorithms in the third round and four, namely BIKE, Classic McEliece, HQC, and SIKE, for further review moved to the fourth round \cite{PQCsurvey}. The L-System based cryptography intends to use one of the standardized algorithms by NIST as potential future work. 

\section{Conclusion}
The leverage on Fractals and L-System algorithms offers an architectural solution ownership of the data and meta data on SMPs and is proposed as a paradigm shift from centralized and columnar databases. It provides the ability to integrate the evaluation of the source to check if it is trusted and implement non-repudiation, which is undeniable traceability to the source. This is an organic approach to solving a major problem in virtual reality. This project provides a base framework to investigate many aspects and needs exploration as explained in Future Work section earlier.
The other impacts of SMPs, including habits addiction, presence of offensive businesses, and much more, are beyond the scope of this term project. In the proposed research, there is the possibility to indirectly solve these impacts of SMPs. As expected, this is highly dependent on the design architecture and other infrastructure, and needs to be investigated in the future. SMPs started with the goal of bringing people together under a platform and later moved on to monetizing it at the expense of their platform users. It calls for global legislation and brainstorming from global communities, and we look forward to that debate and solutions which might be better than what we proposed. \\

%
%



\begin{thebibliography}{99}
\label{references}

\bibitem[Ari22]{alfaidi2022right} Alfaidi, Arij, and Sudhanshu Semwal. "The right to be forgotten: Privacy and security in blockchain with multi-authority based chameleon hash function map-abch solution." In Future of Information and Communication Conference, pp. 861-876. Cham: Springer International Publishing, 2022.

\bibitem[Bar14]{facebookArchitecture} Barrigas, Hugo et al. "Overview of Facebook scalable architecture." In Proceedings of the International Conference on Information Systems and Design of Communication, pp. 173-176. 2014.

\bibitem[Ben82]{mandelbrot1982fractal} Benoit B Mandelbrot. The fractal geometry of nature, volume 1. WH freeman New York, 1982.

\bibitem[Cha24]{chaturvedi2024postQC} Chaturvedi, Atul. "Post-Quantum Cryptography." arXiv preprint arXiv:2402.10576 (2024).

\bibitem[Gla23]{glassner2023QuantumComputing} Glassner, Andrew. "An introduction to quantum computing." In ACM SIGGRAPH 2024 Courses, pp. 1-65. 2024.

\bibitem[Ham21]{smp_UE_shahbaznezhad2021role} Shahbaznezhad, Hamidreza, Rebecca Dolan, and Mona Rashidirad. "The Role of Social Media Content Format and Platform in Users' Engagement Behavior." Journal of Interactive Marketing 53 (2021): 47-65.

\bibitem[Jar18]{JaromeLanierYouTubeHere}
"How we need to remake the internet" | Jaron Lanier, TED, https://www.youtube.com/watch?v=qQ-PUXPVlos  May 2018.

\bibitem[Jud24]{CSAM}  Durbin Introduces Stop CSAM Act to Crack Down on the Proliferation of Child Sexual Abuse Material Online, U.S. Senate Committee on the Judiciary 
https://www.judiciary.senate.gov/press/releases/
durbin-delivers-opening-statement-during-senate-judiciary-committee-hearing-examining-big-techs-failures-to-protect-kids-from-sexual-exploitation-online (January 2024)
 
\bibitem[Kas21]{Recoinnassance} Kashyap, Pavan, and Vinesha Selvarajah. "Analysis of Different Methods of Reconnaissance." In 3rd International Conference on Integrated Intelligent Computing Communication \& Security (ICIIC 2021), pp. 509-519. Atlantis Press, 2021.

\bibitem[Kum20]{PostQuantumCryptography} Kumar, Manoj, and Pratap Pattnaik. "Post quantum cryptography (pqc)-an overview." In 2020 IEEE High Performance Extreme Computing Conference (HPEC), pp. 1-9. IEEE, 2020.
  
\bibitem[Leh15]{elasticity-scalability} Lehrig, Sebastian, Hendrik Eikerling, and Steffen Becker. "Scalability, elasticity, and efficiency in cloud computing: A systematic literature review of definitions and metrics." In Proceedings of the 11th international ACM SIGSOFT conference on quality of software architectures, pp. 83-92. 2015.

\bibitem[McC00]{non-repudiation} McCullagh, Adrian, and William Caelli. "Non-repudiation in the digital environment." (2000).

\bibitem[Pru12]{prusinkiewicz2012algorithmic} Prusinkiewicz, Przemyslaw, and Aristid Lindenmayer. The algorithmic beauty of plants. Springer Science \& Business Media, 2012.

\bibitem[Sha24]{PQCsurvey} Shajahan, Rasha, Kurunandan Jain, and Prabhakar Krishnan. "A Survey on NIST 3 rd Round Post Quantum Digital Signature Algorithms." In 2024 5th International Conference on Mobile Computing and Sustainable Informatics (ICMCSI), pp. 132-140. IEEE, 2024.

\bibitem[Spi02]{ScriptKiddie} Spitzner, Lance. "Know your enemy." Addison-Wesley (2002).

\bibitem[Sto22a]{ScriptKiddieOrNationStates_gaining} Stoddart, Kristan. "Gaining access: attack and defense methods and legacy systems." In Cyberwarfare: Threats to critical infrastructure, pp. 227-280. Cham: Springer International Publishing, 2022.

\bibitem[Sto22b]{HAckingtheHuman} Stoddart, Kristan. "Hacking the Human." In Cyberwarfare: Threats to Critical Infrastructure, pp. 281-349. Cham: Springer International Publishing, 2022.

\bibitem[Tru20]{AI_SFS} Truong, Thanh Cong et al. "Artificial intelligence and cybersecurity: Past, presence, and future." In Artificial intelligence and evolutionary computations in engineering systems, pp. 351-363. Springer Singapore, 2020.

\bibitem[Von1906]{von1906methode} Von Koch, Helge. "On a continuous curve without tangents constructive from elementary geometry." In Classics on fractals, pp. 24-45. CRC Press, 2019.

\bibitem[War19]{ward2019journalism} Ward, Stephen JA. "Journalism ethics." In The handbook of journalism studies, pp. 307-323. Routledge, 2019.

\end{thebibliography}
\end{document}